\documentclass[prd,aps,twocolumn,nofootinbib]{revtex4}

\usepackage[dvipdfm]{graphicx}
\usepackage{amssymb}
\usepackage{epsfig}
\newcommand{\be}{\begin{equation}}
\newcommand{\ee}{\end{equation}}
\newcommand{\ba}{\begin{eqnarray}}
\newcommand{\ea}{\end{eqnarray}}
\newcommand{\bd}{\begin{displaymath}}
\newcommand{\ed}{\end{displaymath}}

\newcommand{\la}{\lambda}

\newcommand{\oa}{\omega}

\newcommand{\La}{\Lambda}

\newcommand{\cF}{{\cal F}}

\newcommand{\p}{\partial}

\newcommand{\ra}{\rightarrow}

\newcommand{\LF}{\left(}
\newcommand{\RF}{\right)}

\newcommand{\Rd}{\right.}

\newcommand{\2}{\frac{1}{2}}

\newcommand{\mx}{\mbox}

\newcommand{\mand}{\mx{ and }}

\newcommand{\Sf}{\varsigma}
\newcommand{\Ff}{\chi}
\newcommand{\Tp}{T_2}

\def\thalf{{\textstyle{\frac{1}{2}}}}

\begin{document}

\title{Thermal Duality and Hagedorn Transition from $p$-adic Strings}

\author{Tirthabir Biswas$^1$}
\author{Jose A. R. Cembranos$^{2,3}$}
\author{Joseph I. Kapusta$^3$}

\affiliation{
{\it $^1$Department of Physics,
St. Cloud State University, St. Cloud, MN 56301}\\
{\it $^2$William I. Fine Theoretical Physics Institute,
University of Minnesota, Minneapolis, MN 55455}\\
{\it $^3$School of Physics and Astronomy,
University of Minnesota, Minneapolis, MN 55455}
}

\date{\today}

\begin{abstract}
We develop the finite temperature theory of $p$-adic string models. We find that the thermal properties of these non-local field theories can be interpreted either as contributions of standard thermal modes with energies proportional to the temperature, or inverse thermal modes with energies proportional to the inverse of the temperature, leading to a {\it thermal duality} at leading order (genus one) analogous to the well known T-duality of string theory. The $p$-adic strings also recover the asymptotic limits (high and low temperature) for arbitrary genus that purely stringy calculations have yielded. We also discuss our findings surrounding the nature of the Hagedorn transition.
\end{abstract}

\maketitle

One of the most interesting thermal features of string theory is the existence of the Hagedorn phase where at high temperatures the energy is not dominated by the massless modes but rather by the most massive string states, leading to a pressureless fluid \cite{jain,vafa,robert}. In fact, a canonical description of the thermal phase indicated a limiting Hagedorn temperature \cite{jain}. Later, however, it was argued that the limiting temperature only corresponds to the emergence of a thermal tachyonic mode making the description of the system in terms of fundamental string excitations invalid \cite{bala-old}. It was further argued that at temperatures larger than the Hagedorn temperature the free energy $\cF$ grows much more slowly, $\cF\propto T^2$, as compared to conventional field theories where $\cF\propto T^4$, and thus the system represents many fewer degrees of freedom than one would have expected from the zero-temperature string spectrum, or even {\it particle} field theories \cite{atick}.

\begin{figure}
\vspace*{0cm}
\centerline{\mbox{\epsfysize=2 cm\epsfxsize=2 cm\epsfbox{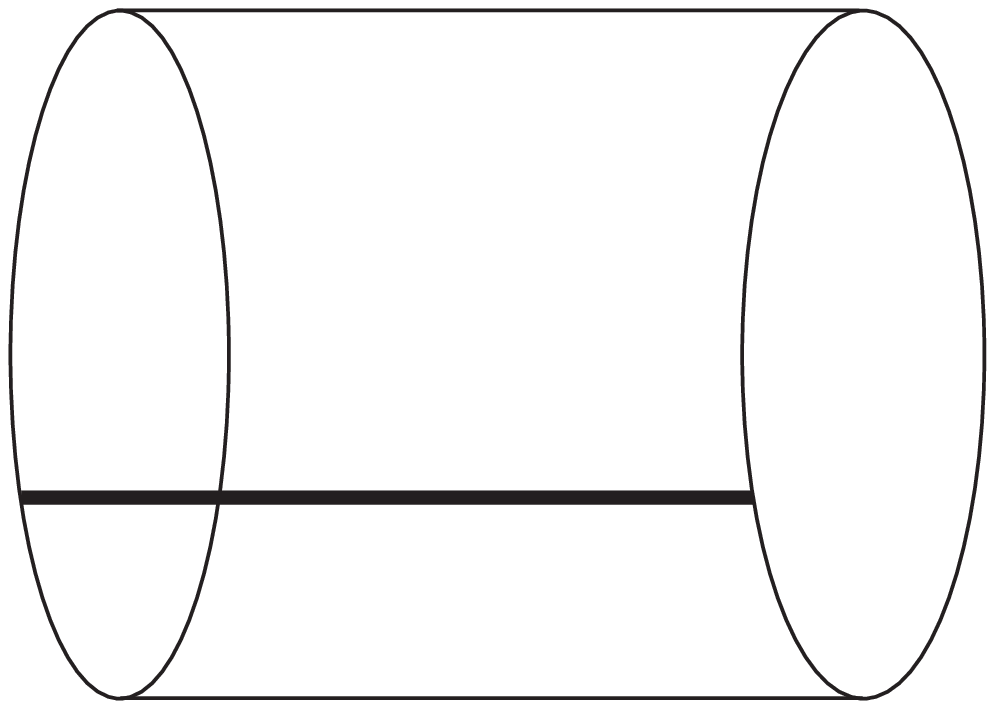}}
\hspace*{0.5 cm}
\mbox{\epsfysize=2 cm\epsfxsize=2 cm\epsfbox{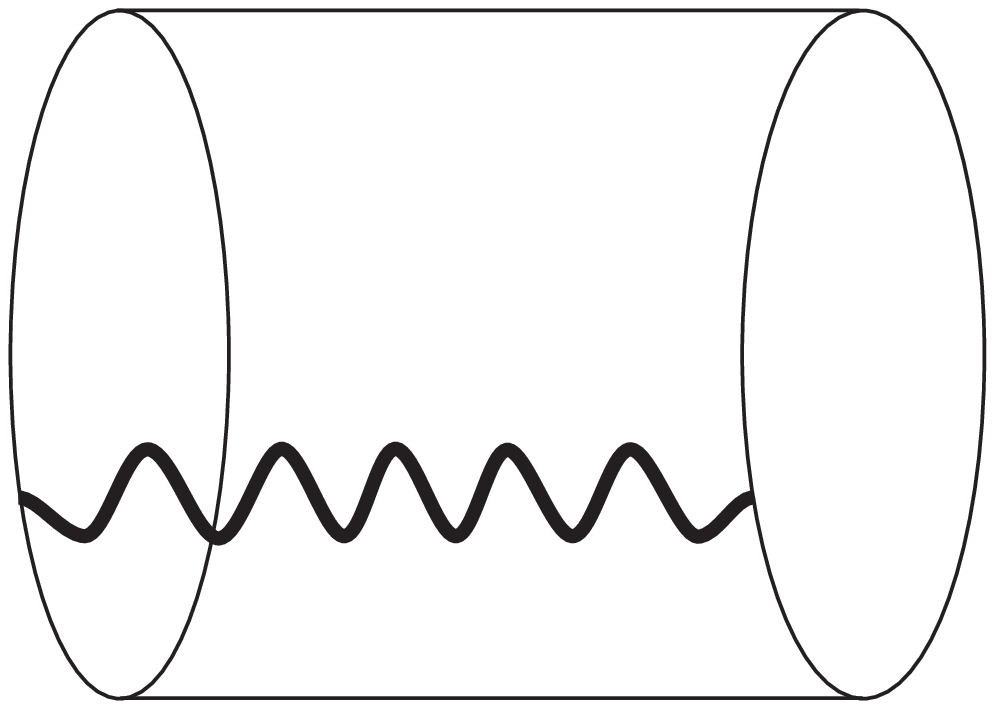}}
\hspace*{0.5 cm}
\mbox{\epsfysize=2 cm\epsfxsize=2 cm\epsfbox{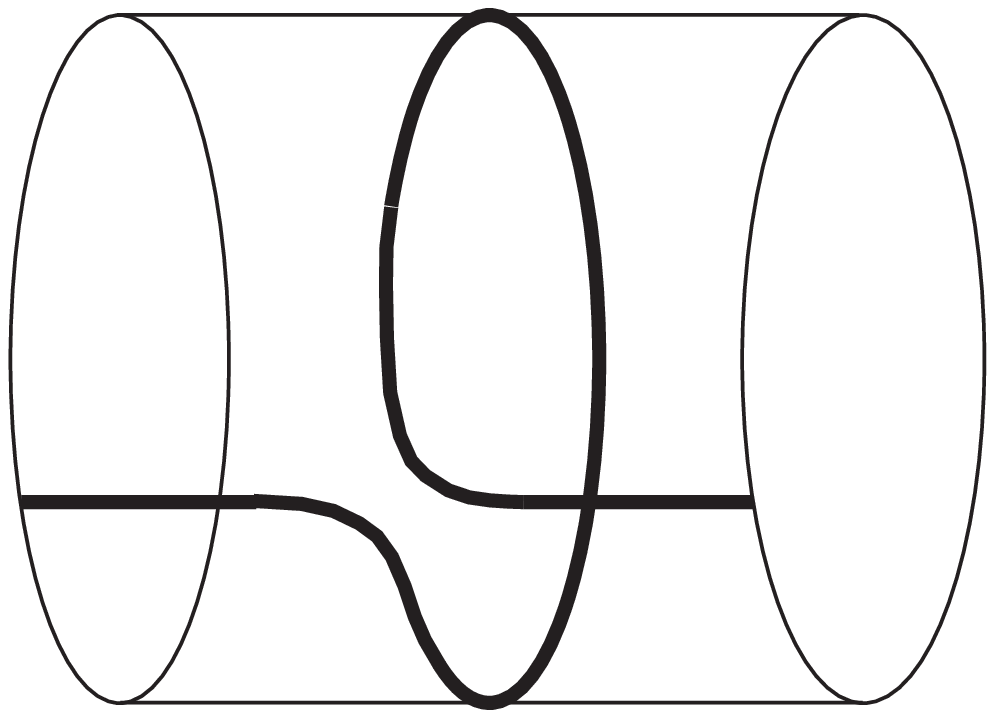}}}
\caption{\footnotesize{Schematic description of string thermal modes. In string theory, the thermal computation can be understood by studying the propagation of fields in $R^{d-1}\times S^1$. The left plot shows the zero thermal state. In string theories, there are not only the standard thermal modes associated with the quantization of momentum, as the one in the middle ($n=11$), but also topological winding modes as the right one ($n_W=1$).}}
\label{smodes}
\end{figure}

Another non-trivial aspect of the stringy partition function that has been studied in the literature \cite{Polchinski:1985zf,atick,bala,dienes} has to do with the (non-)existence of a thermal duality ($T\leftrightarrow 1/T$) in close  analogy with T-duality. In the canonical thermal computation of string models, due to the compact nature of one dimension, there is not only the standard contribution of Matsubara thermal modes: $\omega_n=2\pi n T$ (where $n\in \bf{Z}$); but also the topological contribution of strings wrapped on the torus $S^1$ of circumference $1/T$ \cite{Polchinski:1985zf,atick}: $\omega_W=|n_W| m_s^2/\pi T$, where $m_s$ is at the string scale ($m_s^2=1/2 \alpha'$) and where the winding states are labeled by the first homotopy group $\pi_1(S^1)=\bf{Z}$. See Fig. \ref{smodes}.  Therefore, the thermal duality is defined by making the replacements
\be
2\pi T\longleftrightarrow\frac{m_s^2}{\pi T} \, \mand
n\longleftrightarrow n_W \,.
\ee
This symmetry naively suggests that the partition function verifies
\be
Z(T)=Z\left(\frac{{T_c}^2}{T}\right) \,,
\label{duality}
\ee
with $T_c=m_s/\pi\sqrt{2}$. The existence of such a duality has indeed been verified in several stringy computations, such as in \cite{Polchinski:1985zf,atick,bala}.  Indeed, the Hagedorn temperature is closely related to this critical temperature,
$T_H=T_c/a$, although the exact relation between the temperatures depends on the particular string theory ($a=2$ for the bosonic string, $a=\sqrt{2}$ for the Type II superstring, or $a=1+1/\sqrt{2}$ for the heterotic string \cite{dienes}).

In this paper, we develop a new approach to these questions by studying the thermodynamics of $p$-adic string models that are given by the action \cite{witten,Frampton}
\be
S = \frac{m_s^D}{g_p^2} \int d^Dx \left( -\2 \phi\, p^{-\Box /2m_s^2} \phi
+ \frac{1}{p+1} \phi^{p+1} \right) \,,
\label{action}
\ee
where $\Box = -\partial_t^2 + \nabla_{D-1}^2$ in flat space, and we have defined the p-adic coupling $g_p^2 = g_o^2 (p-1)/p^2$ in terms of the open string coupling $g_o$.  The dimensionless scalar field $\phi(x)$ describes the open string tachyon.  This action was originally derived for $p$ being a prime number, although it has been continued to other positive values and even when $p\ra 1$ \cite{p=1}, corresponding to the bosonic string limit.

The action (\ref{action}) describes a non-local theory for the tachyon field that reproduces the N-point tree amplitudes of non-archimedean open strings \cite{witten,Frampton}. In this sense, it can be understood as a simplified model of the bosonic string which reproduces some aspects of a more realistic theory.  That being said, there are several nontrivial similarities between
$p$-adic string theory and the full string theory. For example, near the true vacuum of the theory, $\phi=0$, the field has no obvious particle-like excitations since its mass-squared goes to infinity.  This is the $p$-adic version of the statement that there are no open string excitations of the tachyon vacuum.  A second similarity is the existence of lump-like soliton solutions representing $p$-adic D-branes \cite{Sen:p-adic}.  The theory of small fluctuations about these lump solutions has a spectrum of equally spaced masses squared \cite{Sen:p-adic,Minahan:ModeInteractions}, just as in the case of normal bosonic string theory. One also obtains a very similar action with exponential kinetic operators (and usually assumed to have a cubic or quartic potential) while quantizing strings on a random lattice \cite{random}.  These field theories are also known to reproduce several features, such as the Regge behaviour \cite{marc}, of their stringy duals.

Like most higher derivative theories, these theories have better ultra-violet (UV) convergence.  Unlike {\it finite-order} higher derivative theories, by virtue of having an infinite set of higher derivative terms, they have been conjectured to be free of ghosts\footnote{There are no perturbative states or poles in the propagator. There are instabilities causing oscillations to grow unboundedly \cite{zwiebach}, but this is due to the presence of the tachyon \cite{futuret}.} and to have a well-posed initial value problem \cite{math,zwiebach,neil} making them phenomenologically interesting to study. In particular, these models have been found to provide novel cosmological properties such as non-slow-roll inflation \cite{bbc}, crossing of the
phantom-divide in the context of dark energy \cite{cosmo}, and for obtaining non-singular bouncing solutions \cite{bouncing} (also see \cite{warren} for similar work with non-local gravitational actions). However, most of these analysis have largely been classical, and little attention has been paid to quantum loop calculations (see \cite{marc,minahan} for an exception).

Here we focus our attention on finite temperature loop calculations. We compare our results with standard thermal properties of string theory and, in particular, with the thermal duality and the stringy Hagedorn phase.  These have found several cosmological applications, especially in the context of cyclic/bouncing cosmologies \cite{cyclic} and thermal structure formation scenarios \cite{thermal}. We will provide results for $D=4$, and $p=3$, but most of the techniques developed here can be generalized to similar non-local models, and similar results are obtained for $D\neq4$, $p\neq 3$ \cite{futurebck}.

The finite temperature action is
\be
S = \int_0^{\beta} d\tau \int d^3x \left[ -\thalf \varphi \,
{\rm e}^{-(\partial^2_{\tau} + \nabla^2)/M^2} \varphi
- \lambda \varphi^4 \right] \,.
\ee
where we have performed the rescaling
\be
\varphi\equiv{m_s^{2} \over g_3}\phi \,,\  \la\equiv -{1\over 18} {g_o^2\over m_s^{4}}\,,\ M^2\equiv {2m_s^2\over \ln 3} \,.
\ee
We note that $\la$ and the rescaled $\varphi$ have mass dimension $-4$ and 2, respectively.  Analysis of this model follows very closely that of the usual scalar theory at finite temperature \cite{KapGale}, except that one has to replace the usual field theory propagator with the exponential one (details will be included in \cite{futurebck})
\be
\frac{1}{p^2} = \frac{1}{{\bf p}^2+\omega_n^2} \ra {\rm e}^{-p^2/M^2}
= {\rm e}^{-({\bf p}^2 + \omega_n^2)/M^2} \,.
\ee
In this case, the free theory does not give any contribution to the partition function \cite{futurebck}, $\ln Z_1=0$, consistent with the fact that $p$-adic strings have no perturbative excitations.

\begin{figure}
\centerline{\mbox{
\epsfxsize=1.5 cm\epsfbox{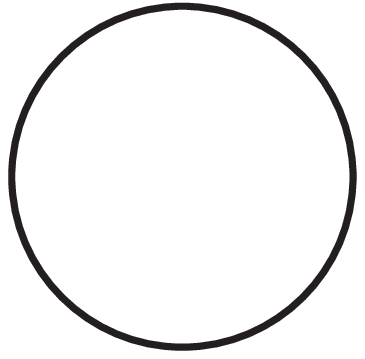}}
\hspace*{2.0 cm}
\mbox{
\epsfxsize=2.5 cm\epsfbox{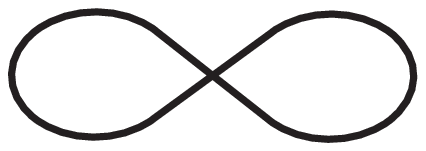}}}
\caption{\footnotesize{The one loop diagram on the left does not contribute to the partition
 function. It shows that the free theory is empty.
 The two loop diagram on the right constitutes the leading order contribution,
 first order in $\lambda$.}}
 \label{leading}
\end{figure}

\begin{figure}[bt]
\begin{center}
\resizebox{7.0cm}{!} {\includegraphics{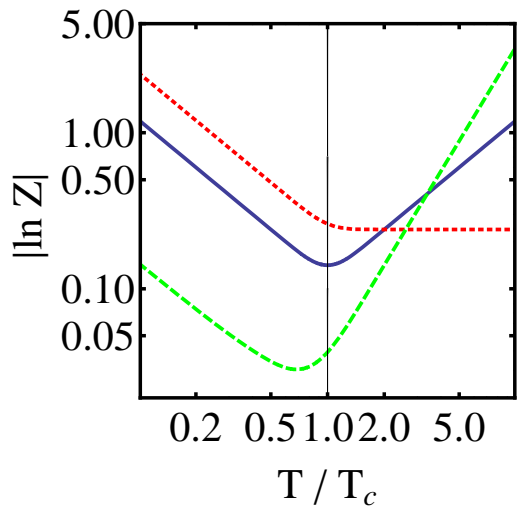}}
\caption{Partition function of the p-adic string model for $p=3$, $D=4$, $T_c=\Tp$, $\lambda T_c^4=-0.04$, and $V=T_c^{-3}$. The symmetry of $Z_2(T)$ (blue straight line) with respect to $T_c$ (central vertical line) shows the exact realization of the thermal duality. However, the duality is broken at high temperatures by higher loop contributions (long dashed green line) or even at low energies by the addition of a self-energy counter-term (short dashed red line) \cite{futurebck}.}
\label{Zfunc}
\end{center}
\end{figure}

Due to the exponential nature of the bare propagator the loop diagrams are expected to be convergent in the ultraviolet limit. The leading contribution comes from the two-loop diagram (see Figs. \ref{leading} and \ref{Zfunc}):
\be
\label{Z2}
\ln Z_2
=-\frac{3\lambda V M^6 T}{2^6 \pi^3}
\Sf^2\left(\frac{2\pi T}{M}\right)\,,
\ee
where the function $\Sf(x)$ can be written in terms of the third Jacobi elliptic theta function: $\Sf(x)=\vartheta_3(0,e^{-x^2})$. Remarkably, one can write this function as:
\be
\label{Sf-duality}
\Sf(x) = \sum_{n=-\infty}^{\infty} {\rm e}^{-n^2x^2}=
\frac{\sqrt{\pi}}{x}
\sum_{m=-\infty}^{\infty} {\rm e}^{-\frac{m^2\pi^2}{x^2}}=
\frac{\sqrt{\pi}}{x}\,\Sf\LF{\pi\over x}\RF \,.
\ee
The first equality shows explicitly the contribution of the $n^{\rm{th}}$ thermal mode. In contrast with a standard quantum field theory, one can see that the higher thermal modes are strongly suppressed at high temperatures; see also Fig. \ref{Zfunc}.  When $x\rightarrow\infty$, the leading term is given by the zero mode ($n=0$). In this limit, the next contribution is given by the first modes ($n=+1$ and $n=-1$), etc. More interestingly, Eq. (\ref{Sf-duality}) suggests that at low temperatures the partition function can be interpreted as the addition of contributions of a different set of modes.  These new modes are not proportional to the temperature, but to the inverse of the temperature. When $x\rightarrow 0$, the leading term is given by the inverse zero mode ($m=0$). The next to leading order contribution is given by the first inverse modes ($m=+1$ and $m=-1$), and so on.

Moreover, we can use the property given by Eq. (\ref{Sf-duality}) to show that Eq. (\ref{duality}) is verified by defining the critical temperature as $T_c=\Tp\equiv M/2\sqrt{\pi}$.  This property is precisely what was predicted by string theory \cite{Polchinski:1985zf,atick}, and it is absolutely
non-trivial in p-adic string models.  As we discussed, this symmetry is present in string theory due to the existence of winding topological string modes,
but in p-adic models there are no obvious topological counterparts because we are working with a quantum field theory.

Since $\Sf(x)$ can be represented as a series of exponentials, one can obtain excellent high and low temperature approximations to $\Sf(x)$. In particular, one finds
\be
\ln Z\ra \left\{\begin{array}{llc} &=-\La V/T\,,& T\ll M \\
&=-4\pi\La T V/ M^2\,,& T\gg M
\end{array}\Rd
\ee
where $\La\equiv 3\lambda \Tp^8$ is the cosmological constant.  This is precisely the kind of asymptotic behavior that has been suggested in the stringy literature \cite{atick,dienes}.

The pressure $P$, entropy density $s$, energy density $\rho$, and the equation of state parameter $\oa$ look like
\ba
P = {\p (T\ln Z)\over \p V}&=& \left\{\begin{array}{lr}-\La\,,& T\ll M \\
-4\pi\La T^2/M^2\,,& T\gg M
\end{array}\Rd\\
s = {\p (T\ln Z)\over V\p T}&=& \left\{\begin{array}{lr}0\,,& T\ll M \\
-8\pi\La T/M^2\,,& T\gg M
\end{array}\Rd\\
\rho = {T^2 \over V}{\p (\ln Z)\over \p T}&=& \left\{\begin{array}{ll}\La\,, & T\ll M \\
-4\pi\La T^2/M^2\,, & T\gg M
\end{array}\Rd\\
\oa\equiv{p\over \rho}
&=& \left\{\begin{array}{rl}-1\,,& T\ll M \\
1\,,& T\gg M \,.
\end{array}\Rd
\ea
The two asymptotic limits mark the two opposite ends of the acceptable range of values for $\oa$. We also note that requiring the entropy density to be positive implies $\la<0$, as it is required for $p$-adic models. This means that the contribution to the cosmological constant is negative, as typical from a stringy approach (however, see \cite{futurebck}).

The next-to-leading order values for the low temperature limit read
\ba
P
&=& -\La-4\La {\rm e}^{-M^2/4T^2} \equiv -\La +P_{\rm H}\,,\\
\rho
&=& \La -2{\La M^2\over T^2}{\rm e}^{-M^2/4T^2}\equiv -\La +\rho_{\rm H} \,.
\ea
The above expressions have a simple interpretation in terms of a cosmological constant plus an effective Hagedorn fluid with energy density and pressure given by $\rho_{\rm H}$ and $P_{\rm H}$, respectively. In particular, we note that the  equation of state parameter of the Hagedorn fluid  vanishes in this limit:
\be
\oa_h\equiv {P_{\rm H}\over \rho_{\rm H}}\approx 2{T^2\over M^2}\ra 0 \mx{ as } {T\over M}\ra 0\ ,
\ee
and the temperature changes very slowly (logarithmically) with the change of energy density.
Qualitatively, this is also what has been argued using stringy computations \cite{jain,vafa,robert}; in the Hagedorn phase the energy density is dominated by the most massive string states and hence behaves as a pressureless fluid, with temperature remaining almost a constant.  However, unlike \cite{jain}, our partition function does not suffer from the problem of having a negative heat capacity, and gives way to the $\oa\approx 1$ stiff fluid phase at high temperatures.

If $g_o^2\ll 1$, we can describe the physics at $T_H$ by using the 2-loop result. From Eq. (\ref{Z2}) and Fig. \ref{Zfunc}, we conclude that our phase transition is smooth and not a first order phase transition, as has been conjectured from string theory. More recently, the possibility of such a smooth transition  has also been suggested~\cite{chaudhuri,dienes}, although in \cite{dienes} a different higher temperature was identified to correspond to  much milder phase transitions in some  supersymmetric string models.

\begin{figure}
\vspace*{0cm}
\centerline{\mbox{
\epsfxsize=3.5 cm\epsfbox{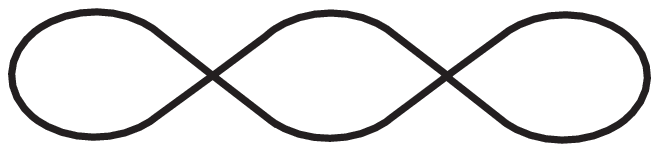}}
\hspace*{1.7 cm}
\mbox{
\epsfxsize=1.7 cm\epsfbox{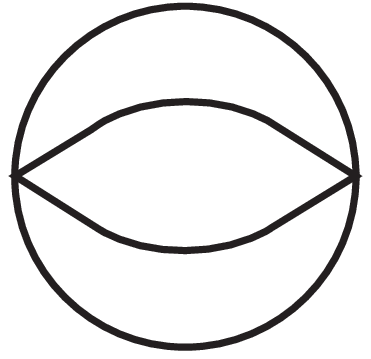}}}
\caption{\footnotesize{Diagrams that contribute at second order in $\lambda$. The one on the left is the three loop contribution to the necklace. The right one is the so called sunset diagram.}}
\label{neckandsun}
\end{figure}

It has been argued that the duality relation (\ref{duality}) must be broken when higher genus interactions are included \cite{atick,bala}. We find the same result.  For example, one can compute the next contribution that is second order in $\lambda$. This contribution is given by the three loop necklace diagram (see Fig. \ref{neckandsun})
\be
\ln Z_{3,\rm neck} =
 9\sqrt{2}\lambda^2 V\,T^2 \Tp^{9} \left[
\Sf\left( \frac{T\sqrt{\pi} }{\Tp}\right) \right]^2
\Sf\left( \frac{T\sqrt{2\pi}}{\Tp}\right)\,,
\ee
together with the sunset diagram (see Fig. \ref{neckandsun})
\be
\ln Z_{\rm sunset} =  \frac{3}{2}\lambda^2 V T^2
\Tp^{9} \,\Ff\left(\frac{T \sqrt{\pi} }{\Tp}\right) \, ,
\ee
where the function $\Ff(x)$ can be written in terms of the third Jacobi theta function.
\bd
\Ff = \int_{-\pi}^{\pi} \frac{d\theta}{2\pi}
\left[ \vartheta_3\left(\thalf\theta,{\rm e}^{-x^2} \right)\right]^4
\ed
\be
\vartheta_3(u,{\rm e}^{-x^2}) \equiv \sum_{n=-\infty}^{\infty} {\rm e}^{-n^2x^2} {\rm e}^{2\,i\,u\,n}
\ee
Therefore, Eq. (\ref{duality}) is only verified at the leading order. In any case, the fact that we can write all the results in terms of $\theta_3(u,{\rm e}^{-x^2})$, which obeys
\be
\vartheta_3(u,{\rm e}^{-x^2})=
\frac{\sqrt{\pi}}{x} {\rm e}^{-u^2/x^2}
\vartheta_3\left(\frac{i\,\pi \,u}{x^2},{\rm e}^{-\pi^2/x^2}\right)\,,
\ee
allows an alternative interpretation in terms of inverse modes, but they need to be weighted in a different way.

For $p$-adic theories, it becomes possible to compute an arbitrary Feynman diagram relatively straightforwardly. Quite remarkably, one finds that at high temperatures $N$-loop diagrams generically give rise to a contribution of the form $\ln Z_N \propto  V m_s^3(g_o^2\,T/m_s)^{N-1}$ \cite{futurebck}, precisely what was argued to be the  contributions at $N$-genus level for open strings \cite{atick}.

In conclusion, we have tried to understand some of the thermal properties of string theory, such as thermal duality, the asymptotic high and low termperature limits, and the Hagedorn transition by studying the 3-adic string model. It is remarkable that this non-local quantum field theory is able to reproduce the fundamental properties of the thermodynamics of string theories. As a cautionary remark, we note that in our analysis we have not included contributions which may arise from solitonic configurations whose effects may be important at high temperatures \cite{futurebck}. Also, the 2-loop result can only be trusted at high temperatures as long as $g_o^2\,T/m_s \ll 1$, whereas at low temperatures it is enough if $g_o^2\ll 1$. A rather interesting feature of these theories is that the free theory is trivial so that all the thermal properties are generated by interactions, even if the model is weakly coupled.

This work was supported by the U.S. DOE Grant Nos. DE-FG02-87ER40328 and DOE/DE-FG02-94ER40823, the FPA 2005-02327 project (DGICYT, Spain), and the CAM/UCM 910309 project.

\end{document}